\documentclass{aa}  
\usepackage{graphics}
\newcommand{\com}[1]{}

\begin{document}
%
\thesaurus{ 3        
(
11.01.2  
11.05.2  
11.17.3  
13.25.2  
13.25.3  
)}
\title{Soft X-ray AGN Luminosity Function from {\it ROSAT} Surveys}
\subtitle{II. Table of the binned Soft X-ray Luminosity Function}
\author{Takamitsu Miyaji  \inst{1,2,3}
\and    G\"unther Hasinger  \inst{3}
\and    Maarten Schmidt     \inst{4}
}

\offprints{T. Miyaji}

\institute{
Department of Physics, Carnegie Mellon University, Pittsburgh, PA 15213, USA
(miyaji@astro.phys.cmu.edu)\thanks{present address for TM}
\and
Max-Planck-Inst. f\"ur Extraterrestrische Physik, Postf. 1603,
D--85740, Garching, Germany 
\and 
Astrophysikalisches Institut Potsdam, An der Sternwarte 16, D--14482
Potsdam, Germany (ghasinger@aip.de)
\and 
California Institute of Technology, Pasadena, CA 91125, USA  
(mxs@deimos.caltech.edu)
}
\date{Received 14 June 2000; accepted }

\titlerunning{{\it ROSAT} AGN Luminosity Function} 
\authorrunning{Miyaji, Hasinger, Schmidt}

\maketitle
\begin{abstract}
%
%
This is the second paper of our investigation of the
0.5-2 keV soft X-ray luminosity function (SXLF) of active galactic
nuclei (AGN)  using results from {\it ROSAT} surveys of various 
depth. The large dynamic range of the combined sample, from shallow large-area
{\it ROSAT} All-Sky Survey (RASS)-based samples to the 
satellite's deepest pointed observation on the Lockman Hole, enabled 
us to trace the behavior of the SXLF. While the first paper (Miyaji, 
Hasinger, Schmidt \cite{paperI}, hereafter paper I)  emphasized  the 
global behavior of the SXLF, cosmological evolution 
and contribution to the soft X-ray background, this paper presents 
actual numerical values for practical use of our results.

 To present the binned SXLF, we have used an improved
estimator, which is free from biases associated with
the conventional $\sum V_{\rm a}^{-1}$ estimator.
 
\end{abstract}

\section{Introduction}
\label{sec:intr}

 The AGN/QSO luminosity function and its evolution with 
cosmic time are key observational quantities for understanding
the origin of and accretion history onto supermassive black holes,
which are now believed to occupy the centers of most galaxies. 
Since X-ray emission is one of the prominent characters of 
the AGN activity, X-ray  surveys are efficient means of sampling 
AGNs for luminosity function and evolution studies.
An X-ray selected sample of AGNs is particularly useful
because optical surveys often use point-like morphology as a
criterion for selecting AGNs (QSOs) among numerous other 
objects, and thus  are likely to miss moderate-luminosity 
intermediate-high redshift AGNs embedded in their host galaxies. 
Also, radio surveys sample only a minor population of AGNs.      
 
The {\it R\"ontgen satellite} ({\it ROSAT})
provided us with soft X-ray surveys with various 
depths, ranging from the {\it ROSAT} All-Sky Survey (RASS
, Voges et al.\cite{rassb}) 
to the {\it ROSAT} Deep Survey (RDS) on the Lockman Hole (Hasinger
et al. \cite{rds1}). Various optical identification programs 
of the survey fields have been conducted and the combination
of these now enabled us to construct the soft X-ray luminosity
function (SXLF) as a function of redshift. 
    
 In paper I., we presented a number of
global representations of the 0.5-2 keV SXLF and investigated
the contribution to the soft X-ray background. 
We showed that our data are not consistent with
the pure-luminosity evolution (PLE), contrary to
the suggestions of a number of previous analyses (e.g. Boyle et al.
\cite{boyle94}; Jones et al. \cite{jones96}).  Instead, 
we find an excess of intermediate-redshift low-luminosity 
AGNs above the PLE case, some sign of which was also recognized 
by Page et al. (\cite{page97}). In view of this, we developed  
two versions of luminosity-dependent density evolution 
(LDDE1 and LDDE2) models, which represent the observed data
very well. An extrapolation of these two LDDE models below the faintest
limit of the survey ($S_{\rm x}=1.7\times 10^{-15}$
$[{\rm erg\,s^{-1}\,cm^{-2}}]$) yields significantly different predictions 
for fainter fluxes, bracketing the range of a possible AGN
contribution to the soft X-ray Background.
Chandra (e.g. Mushotzky et al. \cite{musho00}; 
Hornschemeier et al. \cite{horn00}) and XMM-Newton (\cite{xmmlh1}) 
are probing much fainter sources and spectroscopic identifications
of these will eventually show which of these models is closer to the 
actual behavior of the AGN SXLF. However, because at least some of the 
faint X-ray sources are optically too faint for spectroscopic 
identification even  with the largest telescopes, extending the XLF into 
such faint flux level may be difficult. 
 
 In this second paper, we present practical and convenient 
expressions of the observed SXLF from the {\it ROSAT}
surveys. We present our results mainly for the investigators who  
are interested in particular redshift regimes
and/or comparing their models with observations. 
For this purpose, we show convenient analytical 
expressions in several redshift intervals separately.
These give more accurate representations of the data
in the redshift ranges of interest than those presented
in paper I. We also tabulate the numerical values  of 
the binned SXLF using an improved estimator.
  
We use a Hubble constant $H_0=50\,h_{50}\,$ 
$[{\rm km\,s^{-1}\,Mpc^{-3}}]$.
The $h_{50}$ dependences are explicitly shown. We calculate
the results with common sets of cosmological parameters: 
$(\Omega_{\rm m},\Omega_\Lambda) =$(1.0,0.0), (0.3,0.0). 
and $(\Omega_{\rm m},\Omega_\Lambda) =$ (0.3,0.7).
The symbol ``${\rm Log}$'' represents the base-10 logarithm. 

\section{The summary of the sample}
\label{sec:data}

 We have used soft X-ray sources identified with
AGNs with redshift information from  a combination of {\it ROSAT} 
surveys in various depths/areas from a number of already
published and unpublished sources. Detailed description of 
the definition of the sample, {\it ROSAT} 
countrate-to-flux conversion, and survey area are
shown in paper I. The summary of the samples, which is 
a duplicate of Table 1. of paper I with updated references, is 
shown in Table \ref{tab:surv}. The details of the nature
and completeness of each sample were discussed in paper I.
The limiting flux versus survey area relation were also
shown in paper I.

\begin{table}
\caption[]{{\it ROSAT} Surveys used in the Analysis}
\begin{center}
\begin{tabular}{lccc}
\hline\hline
 Survey$^{\rm a}$ & $S^{\rm lim}_{\rm x14}$ & Area & No. of$^{\rm b}$  \\
       & $10^{-14}[{\rm erg\,s^{-1}\,cm^{-2}}]$ & $[{\rm deg}^2]$ & AGNs  \\
\hline
 RBS   & $\approx 250$ & $2.0 \times 10^{4}$ & 216 \\
 SA-N  & $\approx 13$ &   $685.$  & 130 \\
 RIXOS & $3.0$  &  $15.$   & 205 \\    
 NEP   & $1.0$  & $0.21$  & 13  \\
 UKD   & $0.5$  & $0.16$  & 29  \\
 RDS-Marano & $0.5$ & $0.20$  & 30  \\
 RDS-LH & $0.17-0.9$ & $0.30$ & 68  \\
\hline
\end{tabular}
\end{center}
\label{tab:surv}
$^{\rm a}$ Abbreviations -- RBS: The {\it ROSAT} Bright Survey
(Fischer et al. \cite{rbs1}; Schwope et al. \cite{rbs2}), 
SA-N: The Selected Area-North (Zickgraf et al. \cite{zick97};
Appenzeller et al. \cite{app98}), RIXOS: The {\it ROSAT} International 
X-ray Optical Survey (Mason et al. \cite{rixos}), NEP: The North Ecliptic 
Pole Survey (Bower et al. \cite{nep}); 
UKD: The UK Deep Survey (McHardy et al. \cite{ukd}), 
RDS-Marano: The {\it ROSAT} Deep Survey -- Marano field
(Zamorani et al. \cite{marano},  
RDS-LH: The {\it ROSAT} Deep 
Survey -- Lockman Hole (Hasinger et al. \cite{rds1}; 
Schmidt et al. \cite{rds2}; Lehmann et al. \cite{rds3}). 
$^{\rm b}$ Excluding AGNs with $z<0.015$. 
\end{table}

\section{The SXLF estimation}

  As in paper I, we present the SXLF in the {\it observed} 
0.5-2 keV band, i.e., in the $0.5(1+z)-2(1+z)$
keV range in the object's rest frame. This is equivalent
to assuming an energy index of 1. 
Thus no K-correction was applied for our expressions presented here.
The reasons for this choice are explained in detail in paper I.
This choice is particularly important for this paper, which 
is intended to be used as observational constraints for
population-synthesis-type models (e.g. Madau et al.\cite{madau94};
Comastri et al. \cite{coma95}; Gilli et al. (\cite{gilli99},\cite{gilli01});
Miyaji et al. \cite{miy_cosp}) with various spectral 
assumptions. Because of that, it is more useful to provide 
quantities in a model-independent form rather than 
applying a particular version of model-dependent K-corrections.
By presenting in this manner, one can avoid the difficulty
of reverse K-correcting and re-applying new K-corrections when the 
new results from {\it Chandra} and {\it XMM} provide 
better knowledge of the X-ray spectra of the population.    
 
\subsection{Analytical expressions}
\label{sec:ana}

\begin{table*}
\caption[]{Best-fit parameters for each redshift bin}
\begin{center}
\begin{tabular}{ccrcccccc}
\hline\hline
$z$-range & $z_{\rm c}$ & N & A$^{\rm a}$ & $L_{*}^{\rm a}$ 
  & $\gamma_1$ & $\gamma_2$ & $p$ & KS-prob$^{\rm b}$ \\
\hline 
\multicolumn{9}{c}{$(\Omega_{\rm m},\Omega_\Lambda)\;\;=\;\;(1.0,0.0)$}\\
\hline
0.015-0.2 &  0.1 & 269 &$(2.12\pm0.23)\cdot 10^{-6}$ &
 0.60$^{+0.68}_{-0.32}$ &  0.59$^{+0.23}_{-0.29}$ &
  2.1$^{+0.4}_{-0.3}$ &  4.22$^{+2.53}_{-2.61}$ & 0.99,0.68,0.61\\
0.2-0.4 &  0.3 & 113 &$(3.72\pm0.63)\cdot 10^{-6}$ &
 0.89$^{+1.20}_{-0.46}$ &  0.67$^{+0.30}_{-0.38}$ &
  2.5$^{+0.4}_{-0.3}$ &  5.25$^{+3.48}_{-3.51}$ & 0.93,0.56,0.72\\
0.4-0.8 &  0.6 &  99 &$(1.55\pm0.28)\cdot 10^{-5}$ &
 0.54$^{+0.85}_{-0.29}$ &  0.33$^{+0.52}_{-0.87}$ &
  2.2$^{+0.3}_{-0.2}$ &  5.95$^{+2.29}_{-2.29}$ & 0.85,0.85,0.57\\
0.8-1.6 &  1.2 & 135 &$(1.75\pm0.27)\cdot 10^{-5}$ &
 1.48$^{+1.14}_{-0.56}$ &  0.40$^{+0.41}_{-0.53}$ &
  2.4$^{+0.2}_{-0.2}$ &  4.07$^{+1.33}_{-1.34}$ & 0.99,0.96,0.83\\
1.6-2.3 &  2.2 &  44 &$(3.75\pm1.02)\cdot 10^{-5}$ &
 1.2(*) &  0.0(*) &
  2.1$^{+0.2}_{-0.1}$ & 0(*) & 0.27,0.36,0.19\\
2.3-4.6 &  3.0 &  25 &$(3.93\pm1.41)\cdot 10^{-5}$ &
 1.0(*) & \ldots &
  1.9$^{+0.2}_{-0.2}$ & 0(*) & 0.72,0.99,0.64\\
\hline
\multicolumn{9}{c}{$(\Omega_{\rm m},\Omega_\Lambda)\;\;=\;\;(0.3,0.0)$}\\
\hline
0.015-0.2 &  0.1 & 269 &$(2.07\pm0.23)\cdot 10^{-6}$ &
 0.59$^{+0.71}_{-0.32}$ &  0.59$^{+0.23}_{-0.30}$ &
  2.1$^{+0.4}_{-0.3}$ &  4.13$^{+2.56}_{-2.63}$ & 0.99,0.60,0.54\\
0.2-0.4 &  0.3 & 113 &$(3.33\pm0.56)\cdot 10^{-6}$ &
 0.93$^{+1.30}_{-0.49}$ &  0.67$^{+0.31}_{-0.39}$ &
  2.4$^{+0.4}_{-0.3}$ &  5.31$^{+3.49}_{-3.51}$ & 0.99,0.53,0.79\\
0.4-0.8 &  0.6 &  99 &$(1.03\pm0.19)\cdot 10^{-5}$ &
 0.69$^{+1.07}_{-0.36}$ &  0.37$^{+0.50}_{-0.80}$ &
  2.3$^{+0.3}_{-0.2}$ &  5.90$^{+2.28}_{-2.28}$ & 0.97,0.81,0.57\\
0.8-1.6 &  1.2 & 135 &$(9.28\pm1.44)\cdot 10^{-6}$ &
 2.14$^{+1.67}_{-0.83}$ &  0.42$^{+0.40}_{-0.52}$ &
  2.4$^{+0.2}_{-0.2}$ &  4.13$^{+1.34}_{-1.34}$ & 0.98,0.97,0.72\\
1.6-2.3 &  2.2 &  44 &$(1.90\pm0.51)\cdot 10^{-5}$ &
 1.8(*) &  0.0(*) &
  2.1$^{+0.2}_{-0.1}$ & 0(*) & 0.19,0.41,0.15\\
2.3-4.6 &  3.0 &  25 &$(5.00\pm1.80)\cdot 10^{-5}$ &
 1.0(*) & \ldots &
  1.9$^{+0.2}_{-0.2}$ & 0(*) & 0.64,0.96,0.75\\
\hline
\multicolumn{9}{c}{$(\Omega_{\rm m},\Omega_\Lambda)\;\;=\;\;(0.3,0.7)$}\\
\hline
0.015-0.2 &  0.1 & 269 &$(1.58\pm0.17)\cdot 10^{-6}$ &
 0.71$^{+0.86}_{-0.39}$ &  0.62$^{+0.22}_{-0.29}$ &
  2.1$^{+0.4}_{-0.3}$ &  3.79$^{+2.56}_{-2.64}$ & 0.99,0.54,0.61\\
0.2-0.4 &  0.3 & 113 &$(2.40\pm0.41)\cdot 10^{-6}$ &
 1.09$^{+1.53}_{-0.58}$ &  0.67$^{+0.30}_{-0.39}$ &
  2.4$^{+0.4}_{-0.3}$ &  4.95$^{+3.49}_{-3.51}$ & 0.97,0.56,0.82\\
0.4-0.8 &  0.6 &  99 &$(6.71\pm1.22)\cdot 10^{-6}$ &
 0.85$^{+1.41}_{-0.46}$ &  0.36$^{+0.51}_{-0.87}$ &
  2.2$^{+0.3}_{-0.2}$ &  5.69$^{+2.28}_{-2.27}$ & 0.96,0.81,0.60\\
0.8-1.6 &  1.2 & 135 &$(5.68\pm0.88)\cdot 10^{-6}$ &
 2.69$^{+2.07}_{-1.04}$ &  0.43$^{+0.39}_{-0.51}$ &
  2.4$^{+0.2}_{-0.2}$ &  4.10$^{+1.33}_{-1.34}$ & 0.98,0.97,0.74\\
1.6-2.3 &  2.2 &  44 &$(1.34\pm0.36)\cdot 10^{-5}$ &
 2.0(*) &  0.0(*) &
  2.1$^{+0.1}_{-0.1}$ & 0(*) & 0.14,0.40,0.16\\
2.3-4.6 &  3.0 &  25 &$(4.25\pm1.53)\cdot 10^{-5}$ &
 1.0(*) & \ldots &
  1.9$^{+0.2}_{-0.2}$ & 0(*) & 0.78,0.98,0.76\\
\hline
\end{tabular}
\label{tab:fit}

\end{center}

 Parameter values which have been fixed during the fit are labeled
by '(*)'.  $^{\rm a}$Units -- A: [$h_{50}^3\;{\rm Mpc^{-3}}$],\,\,
 $L_*$: [$10^{44}\;h_{50}^{-2}{\rm erg\;s^{-1}}$].
$^{\rm b}$ The three values are probabilities in two 
1D-KS test for the distribution, $L_{\rm x}$, 1D-KS test for the  
$z$ distribution and the 2D-KS test for the ($L_{\rm x}$,$z$) space 
respectively.

\end{table*}

 First, we find a smooth analytical
function for each redshift bin using a Maximum-likelihood 
fitting. The absolute goodness of the resulting expression can 
then be tested by one- and two-dimensional Kolgomorov-Smirnov tests 
(hereafter, 1D-KS and 2D-KS tests respectively; Press et al.\cite{numrec}; 
Fasano \& Franceschini \cite{ff_2dks}). See paper I for detailed
description of these methods.  These fittings and 
tests can be applied to unbinned data sets thus are
free from artifacts and biases from binning.

 For an analytical expression, we use the {\em smoothed two-
power-law} formula, as we did in paper I. Here,
we fit the data in narrow redshift bins and thus evolution
in each redshift bin is assumed to be a pure density evolution
form:

\begin{equation}
\frac{{\rm d}\;\Phi\,(L_{\rm x},z) }{{\rm d\;Log}\;L_{\rm x}}
  = {A}\;\left[\left(\frac{L_{\rm x}}{{L_*}}\right)^{{\gamma_1}}
         +\left(\frac{L_{\rm x}}{{L_*}}\right)^{{\gamma_2}}
  	\right]^{-1} \cdot \left(\frac{1+z}{1+z_{\rm c}} \right)^p,
\label{eq:ana}
\end{equation}
where $z_{\rm c}$ is the central redshift of the bin. For the highest
redshift bin where the ``break'' is not apparent, we have used 
a single power-law form by neglecting the first term in the square 
bracket in Eq.\ref{eq:ana}.

 The luminosity range of the fit is from   
Log $\,L_{\rm x}$=41.7 to the maximum available luminosity
in the sample. As shown below and in paper I, the SXLF 
below the minimum luminosity has a significant excess
above the smooth extrapolation. This excess smoothly
connects with the SXLF of the non-AGN population (e.g. Hasinger
et al. \cite{has99}) and the X-ray emission may be significantly
contaminated by non-AGN activities.
 
 The set of parameters which give the best fit for each 
redshift bin are shown in Table \ref{tab:fit} along with
the results of the 1D- and 2D- KS tests (see the notes of the table). 
The parameter errors correspond to a likelihood change of 2.7 
(90\% confidence errors). In any case, Eq. \ref{eq:ana} gives 
a statistically satisfactory expression for all redshift bins. 

\subsection{The $\sum V_{\rm a}^{-1}$ Method}

 The $\sum V_{\rm a}^{-1}$ estimator, which is a generalized
version of the original $\sum V_{\rm max}^{-1}$ estimator
(Schmidt \cite{vmax}) applied to a sample composed
of subsamples of different depths (see paper I; Avni \& Bahcall
\cite{veva}), has been widely used for binned luminosity
functions (LF; we use the acronym LF when the discussion is not limited
to the luminosity function in the X-ray band) in the literature.
 However, as discussed in paper I (see also 
Wisotzki \cite{wis}; Page \& Carrera \cite{page99}), 
using it for a {\rm binned} LF estimator can cause 
significant biases, especially if the bin covers the flux range
where the available solid angle of the survey changes rapidly
as a function of flux. Also, the choice of the location
in a ${\rm Log}\,L_{\rm x}$ bin with a non-negligible width  
at which the data point is plotted significantly changes the impression 
of the plot.  
In Fig. 3 of paper I, however, we plotted the  
$\sum V_{\rm a}^{-1}$ estimates, because of the lack of a 
reasonable alternative at the time of writing that paper, with  
caveats on biases associated with the method.
We note that the estimator can be
used in an {\rm unbinned} manner by considering a set of 
delta-functions weighted by $V_{\rm a}^{-1}$ (or $V_{\rm max}^{-1}$) 
at the positions of sample objects in the luminosity space 
(Schmidt \& Green \cite{schm_gre}), and this method is 
free from biases mentioned above. While this unbinned method is 
a powerful tool to predict, e.g.,  the source counts, it 
does not provide practical means of plotting.
In this paper, we have developed an improved estimator,
which is explained in the next subsection. 
  
\subsection{An improved estimator of the binned SXLF}
\label{sec:imp}
   
\begin{table}
\caption[]{The full binned SXLF values -- 1.} 
\begin{center}
\begin{tabular}{ccrrc}
\hline\hline
\multicolumn{5}{c}{$(\Omega_{\rm m},\Omega_\Lambda)\;=\;(1.0,0.0)$}
\\
$z$ & Log\,{$L_{\rm x}$}$^{\rm a}$ 
  & $N^{\rm obs}$ & $N^{\rm mdl}$ &  
     $\frac{{\rm d \Phi^{\rm n}}}{{\rm d\, Log}\,L_{\rm x}}^{\rm b}$    \\
 (1) & (2) & \multicolumn{1}{c}{(3)} & \multicolumn{1}{c}{(4)} 
 & \multicolumn{1}{c}{(5)} \\
 \hline 
.015-0.2 & 41.30-41.70 &   5(3) &   1.0 
 &  $(2.3^{+1.5}_{-1.0})\cdot 10^{-4}$ \\
.015-0.2 & 41.70-42.40 &   8(2) &   6.3 
 &  $(2.8^{+1.3}_{-1.0})\cdot 10^{-5}$ \\
.015-0.2 & 42.40-43.00 &  23(3) &  26.8 
 &  $(7.7^{+1.9}_{-1.6})\cdot 10^{-6}$ \\
.015-0.2 & 43.00-43.50 &  61(7) &  58.4 
 &  $(3.9^{+0.6}_{-0.5})\cdot 10^{-6}$ \\
.015-0.2 & 43.50-43.80 &  56(3) &  52.6 
 &  $(1.6^{+0.2}_{-0.2})\cdot 10^{-6}$ \\
.015-0.2 & 43.80-44.20 &  55(4) &  65.5 
 &  $(4.1^{+0.6}_{-0.6})\cdot 10^{-7}$ \\
.015-0.2 & 44.20-44.50 &  35(1) &  31.6 
 &  $(1.3^{+0.2}_{-0.2})\cdot 10^{-7}$ \\
.015-0.2 & 44.50-45.00 &  30 &  25.0 
 &  $(2.1^{+0.4}_{-0.4})\cdot 10^{-8}$ \\
.015-0.2 & 45.00-45.70 &   1 &   3.0 
 &  $(3.3^{+6.8}_{-2.7})\cdot 10^{-10}$ \\
\\
0.2-0.4 & 41.70-42.10 &   1 &   0.8 
 &  $(1.1^{+2.3}_{-0.9})\cdot 10^{-4}$ \\
0.2-0.4 & 42.10-42.70 &   3(2)&   4.3 
 &  $(2.8^{+2.5}_{-1.5})\cdot 10^{-5}$ \\
0.2-0.4 & 42.70-43.30 &  18(5)&  15.1 
 &  $(1.9^{+0.5}_{-0.4})\cdot 10^{-5}$ \\
0.2-0.4 & 43.30-43.80 &  27(3)&  26.3 
 &  $(5.9^{+1.3}_{-1.1})\cdot 10^{-6}$ \\
0.2-0.4 & 43.80-44.30 &  25(3)&  29.0 
 &  $(1.1^{+0.3}_{-0.2})\cdot 10^{-6}$ \\
0.2-0.4 & 44.30-44.90 &  26(1)&  23.5 
 &  $(9.6^{+2.2}_{-1.9})\cdot 10^{-8}$ \\
0.2-0.4 & 44.90-45.40 &  13 &  11.3 
 &  $(4.7^{+1.6}_{-1.3})\cdot 10^{-9}$ \\
\\
0.4-0.8 & 42.30-42.70 &   0 &   1.1 
 &  $< 7.9\cdot 10^{-5}$ \\
0.4-0.8 & 42.70-43.30 &  13(7)&  10.8 
 &  $(3.1^{+1.1}_{-0.9})\cdot 10^{-5}$ \\
0.4-0.8 & 43.30-43.60 &  14(4)&  11.7 
 &  $(1.8^{+0.6}_{-0.5})\cdot 10^{-5}$ \\
0.4-0.8 & 43.60-44.20 &  38 &  43.2 
 &  $(3.9^{+0.7}_{-0.6})\cdot 10^{-6}$ \\
0.4-0.8 & 44.20-44.80 &  19(1)&  16.8 
 &  $(3.3^{+0.9}_{-0.7})\cdot 10^{-7}$ \\
0.4-0.8 & 44.80-45.40 &  10 &   8.9 
 &  $(1.5^{+0.6}_{-0.5})\cdot 10^{-8}$ \\
0.4-0.8 & 45.40-46.50 &   5 &   6.5 
 &  $(1.3^{+0.8}_{-0.6})\cdot 10^{-10}$ \\
\\
0.8-1.6 & 42.70-43.30 &   5 &   2.9 
 &  $(8.9^{+5.6}_{-3.8})\cdot 10^{-5}$ \\
0.8-1.6 & 43.30-43.90 &  23 (2)&  26.9 
 &  $(2.4^{+0.6}_{-0.5})\cdot 10^{-5}$ \\
0.8-1.6 & 43.90-44.50 &  55 &  53.0 
 &  $(8.2^{+1.2}_{-1.1})\cdot 10^{-6}$ \\
0.8-1.6 & 44.50-45.10 &  39 &  36.0 
 &  $(5.6^{+1.0}_{-0.9})\cdot 10^{-7}$ \\
0.8-1.6 & 45.10-45.70 &   5 &   9.0 
 &  $(1.1^{+0.7}_{-0.5})\cdot 10^{-8}$ \\
0.8-1.6 & 45.70-46.20 &   4 &   3.1 
 &  $(1.2^{+0.9}_{-0.6})\cdot 10^{-9}$ \\
0.8-1.6 & 46.20-46.90 &   4 &   2.3 
 &  $(6.1^{+4.5}_{-2.9})\cdot 10^{-11}$ \\
\\
1.6-2.3 & 43.60-44.20 &   9 &  13.1 
 &  $(1.8^{+0.8}_{-0.6})\cdot 10^{-5}$ \\
1.6-2.3 & 44.20-44.80 &  18 &  12.9 
 &  $(5.8^{+1.6}_{-1.4})\cdot 10^{-6}$ \\
1.6-2.3 & 44.80-45.50 &  14 &  10.5 
 &  $(2.5^{+0.8}_{-0.7})\cdot 10^{-7}$ \\
1.6-2.3 & 45.50-46.10 &   2 &   2.3 
 &  $(6.7^{+8.0}_{-4.3})\cdot 10^{-9}$ \\
1.6-2.3 & 46.10-46.80 &   0 &   1.5 
 &  $< 4.6\cdot 10^{-10}$ \\
1.6-2.3 & 46.80-47.40 &   1 &   1.0 
 &  $(1.2^{+2.5}_{-1.0})\cdot 10^{-11}$ \\
\\
2.3-4.6 & 43.70-44.10 &   2 &   3.4 
 &  $(3.6^{+4.3}_{-2.3})\cdot 10^{-5}$ \\
2.3-4.6 & 44.10-44.80 &  12 &  10.5 
 &  $(6.0^{+2.2}_{-1.7})\cdot 10^{-6}$ \\
2.3-4.6 & 44.80-45.40 &   7 &   4.3 
 &  $(4.6^{+2.3}_{-1.7})\cdot 10^{-7}$ \\
2.3-4.6 & 45.40-46.20 &   2 &   3.2 
 &  $(7.8^{+9.4}_{-5.0})\cdot 10^{-9}$ \\
2.3-4.6 & 46.20-47.00 &   0 &   1.4 
 &  $< 5.4\cdot 10^{-10}$ \\
2.3-4.6 & 47.00-47.50 &   2 &   1.0 
 &  $(3.8^{+4.6}_{-2.5})\cdot 10^{-11}$ \\
\\
\hline
\end{tabular}
\end{center}
\label{tab:xlf_f10}
Notes:  
$^{\rm a} L_{\rm x}[h_{50}^{-2}\,{\rm erg\,s^{-1}}]$ in 0.5-2 [keV].    
$^{\rm b} [h_{50}^3\,{\rm Mpc}^{-3}]$. $^c$  The AGNs in this row
are outside of the luminosity range used for the model fit.
\end{table}

\begin{table}
\caption[]{The full binned SXLF values -- 2.} 
\begin{center}
\begin{tabular}{ccrrc}
\hline\hline
\multicolumn{5}{c}{$(\Omega_{\rm m},\Omega_\Lambda)\;=\;(0.3,0.0)$}
\\
$z$ & Log\,{$L_{\rm x}$}$^{\rm a}$ 
  & $N^{\rm obs}$ & $N^{\rm mdl}$ &  
     $\frac{{\rm d \Phi^{\rm n}}}{{\rm d\, Log}\,L_{\rm x}}^{\rm b}$    \\
 (1) & (2) & \multicolumn{1}{c}{(3)} & \multicolumn{1}{c}{(4)} 
 & \multicolumn{1}{c}{(5)} \\
 \hline 
.015-0.2 & 41.30-41.70 &   5(3)&   1.0 
 &  $(2.3^{+1.4}_{-1.0})\cdot 10^{-4}$ \\
.015-0.2 & 41.70-42.40 &   7(2)&   6.1 
 &  $(2.5^{+1.2}_{-0.9})\cdot 10^{-5}$ \\
.015-0.2 & 42.40-43.00 &  24(3)&  26.0 
 &  $(8.1^{+1.9}_{-1.6})\cdot 10^{-6}$ \\
.015-0.2 & 43.00-43.50 &  58(6)&  56.4 
 &  $(3.7^{+0.5}_{-0.5})\cdot 10^{-6}$ \\
.015-0.2 & 43.50-43.80 &  55(4)&  50.5 
 &  $(1.6^{+0.2}_{-0.2})\cdot 10^{-6}$ \\
.015-0.2 & 43.80-44.20 &  58(4)&  64.4 
 &  $(4.3^{+0.6}_{-0.6})\cdot 10^{-7}$ \\
.015-0.2 & 44.20-44.60 &  47(1)&  39.8 
 &  $(1.1^{+0.2}_{-0.2})\cdot 10^{-7}$ \\
.015-0.2 & 44.60-45.20 &  19 &  20.6 
 &  $(8.2^{+2.3}_{-1.9})\cdot 10^{-9}$ \\
.015-0.2 & 45.20-45.80 &   1 &   1.3 
 &  $(3.9^{+8.0}_{-3.2})\cdot 10^{-10}$ \\
\\
0.2-0.4 & 41.70-42.10 &   1 &   0.6 
 &  $(1.3^{+2.6}_{-1.0})\cdot 10^{-4}$ \\
0.2-0.4 & 42.10-42.70 &   2(1)&   4.2 
 &  $(1.8^{+2.1}_{-1.1})\cdot 10^{-5}$ \\
0.2-0.4 & 42.70-43.40 &  21(6)&  19.4 
 &  $(1.4^{+0.4}_{-0.3})\cdot 10^{-5}$ \\
0.2-0.4 & 43.40-43.80 &  23(3)&  22.0 
 &  $(5.0^{+1.2}_{-1.0})\cdot 10^{-6}$ \\
0.2-0.4 & 43.80-44.30 &  26(3)&  28.7 
 &  $(1.1^{+0.3}_{-0.2})\cdot 10^{-6}$ \\
0.2-0.4 & 44.30-44.80 &  23(1)&  21.3 
 &  $(1.3^{+0.3}_{-0.3})\cdot 10^{-7}$ \\
0.2-0.4 & 44.80-45.40 &  17 &  14.6 
 &  $(6.8^{+2.0}_{-1.6})\cdot 10^{-9}$ \\
\\
0.4-0.8 & 42.20-42.80 &   0 &   1.4 
 &  $< 5.3\cdot 10^{-5}$ \\
0.4-0.8 & 42.80-43.25 &   8(5)&   7.1 
 &  $(2.3^{+1.0}_{-0.8})\cdot 10^{-5}$ \\
0.4-0.8 & 43.25-43.60 &  13(6)&  10.7 
 &  $(1.5^{+0.5}_{-0.4})\cdot 10^{-5}$ \\
0.4-0.8 & 43.60-44.10 &  34 &  33.4 
 &  $(5.1^{+1.0}_{-0.9})\cdot 10^{-6}$ \\
0.4-0.8 & 44.10-44.80 &  25 &  28.4 
 &  $(3.6^{+0.8}_{-0.7})\cdot 10^{-7}$ \\
0.4-0.8 & 44.80-45.40 &  13(1)&  10.0 
 &  $(1.9^{+0.7}_{-0.5})\cdot 10^{-8}$ \\
0.4-0.8 & 45.40-46.60 &   6 &   7.4 
 &  $(1.1^{+0.6}_{-0.5})\cdot 10^{-10}$ \\
\\
0.8-1.6 & 43.10-43.60 &   6 &   6.3 
 &  $(2.2^{+1.3}_{-0.9})\cdot 10^{-5}$ \\
0.8-1.6 & 43.60-44.10 &  28(2)&  26.1 
 &  $(1.4^{+0.3}_{-0.3})\cdot 10^{-5}$ \\
0.8-1.6 & 44.10-44.60 &  46 &  45.0 
 &  $(4.5^{+0.7}_{-0.7})\cdot 10^{-6}$ \\
0.8-1.6 & 44.60-45.30 &  42 &  40.6 
 &  $(3.0^{+0.5}_{-0.5})\cdot 10^{-7}$ \\
0.8-1.6 & 45.30-45.90 &   5 &   9.0 
 &  $(4.8^{+3.0}_{-2.1})\cdot 10^{-9}$ \\
0.8-1.6 & 45.90-46.50 &   4 &   3.9 
 &  $(3.4^{+2.5}_{-1.6})\cdot 10^{-10}$ \\
0.8-1.6 & 46.50-47.00 &   4 &   1.6 
 &  $(4.0^{+2.9}_{-1.9})\cdot 10^{-11}$ \\
\\
1.6-2.3 & 43.80-44.50 &  13 &  16.7 
 &  $(9.3^{+3.2}_{-2.5})\cdot 10^{-6}$ \\
1.6-2.3 & 44.50-45.10 &  14 &  12.5 
 &  $(1.4^{+0.5}_{-0.4})\cdot 10^{-6}$ \\
1.6-2.3 & 45.10-45.70 &  14 &   8.8 
 &  $(1.1^{+0.4}_{-0.3})\cdot 10^{-7}$ \\
1.6-2.3 & 45.70-46.40 &   2 &   2.9 
 &  $(2.1^{+2.5}_{-1.3})\cdot 10^{-9}$ \\
1.6-2.3 & 46.40-47.00 &   0 &   1.3 
 &  $< 2.3\cdot 10^{-10}$ \\
1.6-2.3 & 47.00-47.60 &   1 &   1.1 
 &  $(6.0^{+ 12}_{-4.9})\cdot 10^{-12}$ \\
\\
2.3-4.6 & 44.10-44.80 &   9 &   9.9 
 &  $(6.2^{+2.7}_{-2.0})\cdot 10^{-6}$ \\
2.3-4.6 & 44.80-45.40 &   6 &   6.1 
 &  $(3.8^{+2.1}_{-1.5})\cdot 10^{-7}$ \\
2.3-4.6 & 45.40-46.00 &   7 &   4.6 
 &  $(4.2^{+2.1}_{-1.5})\cdot 10^{-8}$ \\
2.3-4.6 & 46.00-46.60 &   1 &   1.9 
 &  $(1.0^{+2.2}_{-0.9})\cdot 10^{-9}$ \\
2.3-4.6 & 46.60-47.10 &   0 &   1.2 
 &  $< 3.2\cdot 10^{-10}$ \\
2.3-4.6 & 47.10-47.60 &   2 &   1.0 
 &  $(3.7^{+4.5}_{-2.4})\cdot 10^{-11}$ \\
\\

\hline
\end{tabular}
\end{center}
\label{tab:xlf_o3}
Notes:  
$^{\rm a} L_{\rm x}[h_{50}^{-2}\,{\rm erg\,s^{-1}}]$ in 0.5-2 [keV].    
$^{\rm b} [h_{50}^3\,{\rm Mpc}^{-3}]$. $^c$  The AGNs in this row
are outside of the luminosity range used for the model fit.
\end{table}

\begin{table}
\caption[]{The full binned SXLF values -- 3.} 
\begin{center}
\begin{tabular}{ccrrc}
\hline\hline
\multicolumn{5}{c}{$(\Omega_{\rm m},\Omega_\Lambda)\;=\;(0.3,0.7)$}
\\
$z$ & Log\,{$L_{\rm x}$}$^{\rm a}$ 
  & $N^{\rm obs}$ & $N^{\rm mdl}$ &  
     $\frac{{\rm d \Phi^{\rm n}}}{{\rm d\, Log}\,L_{\rm x}}^{\rm b}$    \\
 (1) & (2) & \multicolumn{1}{c}{(3)} & \multicolumn{1}{c}{(4)} 
 & \multicolumn{1}{c}{(5)} \\
\hline 
.015-0.2 & 41.30-41.70 &   5(3) &   1.0 
 &  $(2.3^{+1.4}_{-1.0})\cdot 10^{-4}$ \\
.015-0.2 & 41.70-42.60 &  12(3)&  10.6 
 &  $(2.0^{+0.7}_{-0.6})\cdot 10^{-5}$ \\
.015-0.2 & 42.60-43.00 &  19(2)&  20.5 
 &  $(6.4^{+1.8}_{-1.5})\cdot 10^{-6}$ \\
.015-0.2 & 43.00-43.50 &  58(6)&  53.5 
 &  $(3.6^{+0.5}_{-0.5})\cdot 10^{-6}$ \\
.015-0.2 & 43.50-43.80 &  49(3)&  49.1 
 &  $(1.4^{+0.2}_{-0.2})\cdot 10^{-6}$ \\
.015-0.2 & 43.80-44.20 &  62(5)&  67.5 
 &  $(4.3^{+0.6}_{-0.6})\cdot 10^{-7}$ \\
.015-0.2 & 44.20-44.60 &  41(1)&  41.8 
 &  $(9.1^{+1.6}_{-1.4})\cdot 10^{-8}$ \\
.015-0.2 & 44.60-45.10 &  27 &  21.9 
 &  $(1.4^{+0.3}_{-0.3})\cdot 10^{-8}$ \\
.015-0.2 & 45.10-45.80 &   1 &   2.5 
 &  $(2.5^{+5.1}_{-2.1})\cdot 10^{-10}$ \\
\\
0.2-0.4 & 41.70-42.20 &   1 &   0.8 
 &  $(7.8^{+ 16}_{-6.5})\cdot 10^{-5}$ \\
0.2-0.4 & 42.20-42.80 &   2(1)&   4.5 
 &  $(1.2^{+1.4}_{-0.8})\cdot 10^{-5}$ \\
0.2-0.4 & 42.80-43.40 &  19(6)&  14.9 
 &  $(1.3^{+0.4}_{-0.3})\cdot 10^{-5}$ \\
0.2-0.4 & 43.40-43.80 &  21(2)&  21.3 
 &  $(4.0^{+1.0}_{-0.9})\cdot 10^{-6}$ \\
0.2-0.4 & 43.80-44.30 &  26(3)&  28.2 
 &  $(1.1^{+0.2}_{-0.2})\cdot 10^{-6}$ \\
0.2-0.4 & 44.30-44.70 &  17(2)&  19.8 
 &  $(1.3^{+0.4}_{-0.3})\cdot 10^{-7}$ \\
0.2-0.4 & 44.70-45.30 &  23 &  16.6 
 &  $(1.5^{+0.4}_{-0.3})\cdot 10^{-8}$ \\
0.2-0.4 & 45.30-46.00 &   4 &   5.7 
 &  $(2.0^{+1.5}_{-1.0})\cdot 10^{-10}$ \\
\\
0.4-0.8 & 42.20-42.80 &   0 &   0.8 
 &  $< 6.8\cdot 10^{-5}$ \\
0.4-0.8 & 42.80-43.25 &   5(3)&   5.5 
 &  $(1.3^{+0.8}_{-0.6})\cdot 10^{-5}$ \\
0.4-0.8 & 43.25-43.60 &  13(7)&   9.8 
 &  $(1.2^{+0.4}_{-0.3})\cdot 10^{-5}$ \\
0.4-0.8 & 43.60-44.10 &  29(1)&  30.3 
 &  $(4.0^{+0.9}_{-0.7})\cdot 10^{-6}$ \\
0.4-0.8 & 44.10-44.80 &  31 &  34.0 
 &  $(3.7^{+0.8}_{-0.7})\cdot 10^{-7}$ \\
0.4-0.8 & 44.80-45.40 &  15(1)&  11.1 
 &  $(2.1^{+0.7}_{-0.5})\cdot 10^{-8}$ \\
0.4-0.8 & 45.40-46.00 &   5 &   5.8 
 &  $(6.1^{+3.8}_{-2.6})\cdot 10^{-10}$ \\
0.4-0.8 & 46.00-46.61 &   1 &   2.5 
 &  $(1.2^{+2.6}_{-1.0})\cdot 10^{-11}$ \\
\\
0.8-1.6 & 43.10-43.60 &   6 &   3.9 
 &  $(2.5^{+1.4}_{-1.0})\cdot 10^{-5}$ \\
0.8-1.6 & 43.60-44.10 &  19(2)&  22.5 
 &  $(7.9^{+2.2}_{-1.8})\cdot 10^{-6}$ \\
0.8-1.6 & 44.10-44.60 &  44 &  41.4 
 &  $(3.8^{+0.7}_{-0.6})\cdot 10^{-6}$ \\
0.8-1.6 & 44.60-45.30 &  51 &  47.7 
 &  $(3.2^{+0.5}_{-0.4})\cdot 10^{-7}$ \\
0.8-1.6 & 45.30-46.00 &   7 &  10.7 
 &  $(4.5^{+2.3}_{-1.7})\cdot 10^{-9}$ \\
0.8-1.6 & 46.00-46.50 &   4 &   3.2 
 &  $(3.2^{+2.3}_{-1.5})\cdot 10^{-10}$ \\
0.8-1.6 & 46.50-47.00 &   4 &   1.9 
 &  $(3.5^{+2.6}_{-1.7})\cdot 10^{-11}$ \\
\\
1.6-2.3 & 43.80-44.50 &   9 &  14.4 
 &  $(5.6^{+2.4}_{-1.8})\cdot 10^{-6}$ \\
1.6-2.3 & 44.50-45.20 &  18 &  15.0 
 &  $(1.1^{+0.3}_{-0.3})\cdot 10^{-6}$ \\
1.6-2.3 & 45.20-45.90 &  14 &   8.9 
 &  $(5.2^{+1.7}_{-1.4})\cdot 10^{-8}$ \\
1.6-2.3 & 45.90-46.60 &   2 &   3.0 
 &  $(7.7^{+9.2}_{-4.9})\cdot 10^{-10}$ \\
1.6-2.3 & 46.60-47.00 &   0 &   0.6 
 &  $< 2.9\cdot 10^{-10}$ \\
1.6-2.3 & 47.00-47.63 &   1 &   1.6 
 &  $(4.4^{+9.1}_{-3.6})\cdot 10^{-12}$ \\
\\
2.3-4.6 & 44.10-44.80 &   7 &   9.0 
 &  $(4.6^{+2.3}_{-1.7})\cdot 10^{-6}$ \\
2.3-4.6 & 44.80-45.40 &   8 &   6.6 
 &  $(4.1^{+1.9}_{-1.4})\cdot 10^{-7}$ \\
2.3-4.6 & 45.40-46.20 &   7 &   5.8 
 &  $(1.9^{+0.9}_{-0.7})\cdot 10^{-8}$ \\
2.3-4.6 & 46.20-47.00 &   1 &   2.2 
 &  $(2.1^{+4.4}_{-1.7})\cdot 10^{-10}$ \\
2.3-4.6 & 47.00-47.63 &   2 &   1.3 
 &  $(3.0^{+3.7}_{-2.0})\cdot 10^{-11}$ \\
\\
\hline
\end{tabular}
\end{center}
\label{tab:xlf_f3}
Notes:  
$^{\rm a} L_{\rm x}[h_{50}^{-2}\,{\rm erg\,s^{-1}}]$ in 0.5-2 [keV].    
$^{\rm b} [h_{50}^3\,{\rm Mpc}^{-3}]$. $^c$  The AGNs in this row
are outside of the luminosity range used for the model fit.
\end{table}

\begin{figure}
\resizebox{\hsize}{!}{\includegraphics{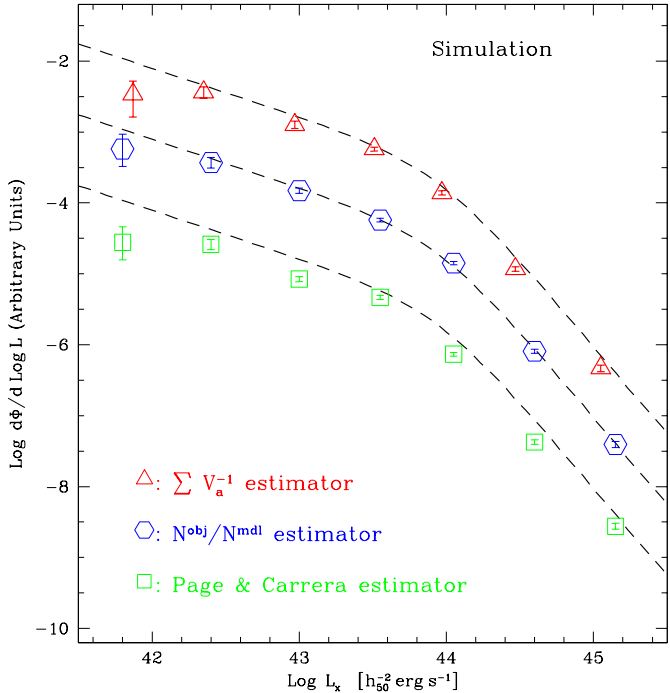}}
\caption[]{The binned XLFs from a simulated sample using
three different estimators (symbols with error bars as labeled) are 
compared with the underlying ``true'' XLF represented by dashed lines. 
The vertical positions of the three different estimators have
been shifted vertically for display.} 
\label{fig:compest}
\end{figure}

As an alternative to the $\sum V_{\rm a}^{-1}$ method,
we have developed the following estimator for the binned
LF, which is free from most biases unavoidable
in the $\sum V_{\rm a}^{-1}$ method. In Sect. \ref{sec:ana}, we have 
found a smooth analytical function which describes the behavior of the
SXLF in a given redshift range. Having the best-fit smooth function,
the estimated numerical value for the SXLF in a given
bin in the $(L_{\rm x},z)$-space is:
\begin{equation}
\frac{{\rm d \Phi^{\rm n}}}{\rm d\, Log\,L_{\rm x}}(L_{{\rm x}i},z_i)
=\frac{{\rm d \Phi^{\rm mdl}}}{\rm d\, Log\,L_{\rm x}}(L_{{\rm x}i},z_i)
\cdot \frac{N^{\rm obs}_i}{N^{\rm mdl}_i},
\label{eq:est}
\end{equation}
where $L_{{\rm x}i}$ and $z_i$ are the luminosity and redshift
representative of the $i$-th bin, $\frac{{\rm d \Phi^{\rm mdl}}}
{\rm d\, Log\,L_{\rm x}}$ is the best-fit analytical expression
evaluated at this point, $N^{\rm obs}_i$ is the actual number 
of AGNs observed in the $i$-th bin, and $N^{\rm mdl}_i$ is
the predicted number of AGNs in the bin from the best-fit
analytical expression. Hereafter, we refer to  Eq. \ref{eq:est} as
the ``$N^{\rm obs}/N^{\rm mdl}$ estimator''. Note that the estimator 
proposed by Page \& Carrera (\cite{page99}) (hereafter PC) is a special 
case of  Eq. \ref{eq:est} where 
$\frac{{\rm d \Phi^{\rm mdl}}}{\rm d\,L_{\rm x}}\propto {\rm const}$
(or  $\frac{{\rm d} \Phi^{\rm mdl}}{{\rm d\,Log}\, L_{\rm x}}\propto 
L_{\rm x}$).

Another advantage of this estimator over $\sim V_{\rm a}^{-1}$ is that 
exact errors at a given significance can be evaluated using  
Poisson statistics. One disadvantage of this estimator is that it 
is model-dependent, at least in principle. Since our analytical 
expressions are satisfactory representations in any case and the 
estimator is not sensitive to the details of the underlying model, the 
uncertainties due to the model dependence are practically negligible. 

 In order to compare the goodness of the estimators,
we performed simulations. Using the actual best-fit model
for the $0.2<z<0.4$ bin for 
$(\Omega_{\rm m},\Omega_\Lambda)\;\;=\;\;(1.0,0.0)$, 
we generated a set of simulated AGNs. The number of 
simulated AGNs are 10 times those of the actual sample in order 
to reduce the Poisson errors. 
  Using the simulated AGNs and the actual flux-area relation of our 
combined sample, we estimated binned SXLFs using three
different estimators: $\sum V_{\rm a}^{-1}$, $N^{\rm obs}/N^{\rm mdl}$ 
(Eq. \ref{eq:est}) and that of PC. 
The results are compared with the underlying SXLF, which was used to 
generate the simulated AGNs, in Fig. \ref{fig:compest}. For the models 
to evaluate $N^{\rm mdl}$, we used the re-fitted model using the 
simulated sample rather than the original model. The 
1$\sigma$ errors for the $N^{\rm obs}/N^{\rm mdl}$ (Eq. \ref{eq:est}) and the 
PC estimators are Poisson errors calculated
using Eqs. (7) and (12) of Gehrels (\cite{gehrels}). On the other hand, 
the errors for the $\sum V_{\rm a}^{-1}$ estimator are from Eq. (3) of 
paper I and are inaccurate for bins with a small number of AGNs.

 As shown in Fig. \ref{fig:compest}, the $N^{\rm obs}/N^{\rm mdl}$
estimator best represents the original model and no estimated
point deviates from the underlying model by more than 2$\sigma$.
The $\sum V_{\rm a}^{-1}$ estimator underestimates the XLF in
the lowest luminosity bin as found in PC.
We note that the PC estimator also systematically 
underestimates the LF in this particular case of the underlying
model and the flux-area relation. 
This is expected because their estimator implicitly builds in the assumption 
$\frac{{\rm d \Phi^{\rm mdl}}}{\rm d\,L_{\rm x}}\propto {\rm const}$
as the underlying LF shape. This is much more weighted towards 
higher luminosities than any part of the realistic AGN XLF.
 Since the amount of this bias depends on the underlying model 
and the flux-area relation, as well as the points in the bin where 
the data are plotted, it is not surprising that the bias is not 
apparent in Fig. 2 of PC. They have also compensated for this bias upon 
comparing the estimated LF with 
a model. Instead of correcting the estimated binned LF using a good 
model (which our $N^{\rm obs}/N^{\rm mdl}$ estimator does), they calculated 
the ``model-expectated value of the estimator'' to compare
with the estimated value from the data.      
Detailed investigation and comparison of these different
estimators in various cases are beyond the scope of this paper. 
Judging from this simulation, the above discussion on biases, and that 
the exact Poisson errors can be used for errors, we choose 
to use the $N^{\rm obs}/N^{\rm mdl}$ estimator for our plots and tabulation.
     
\subsection{The binned SXLF results}

\begin{figure*}
\resizebox{\hsize}{!}{\includegraphics{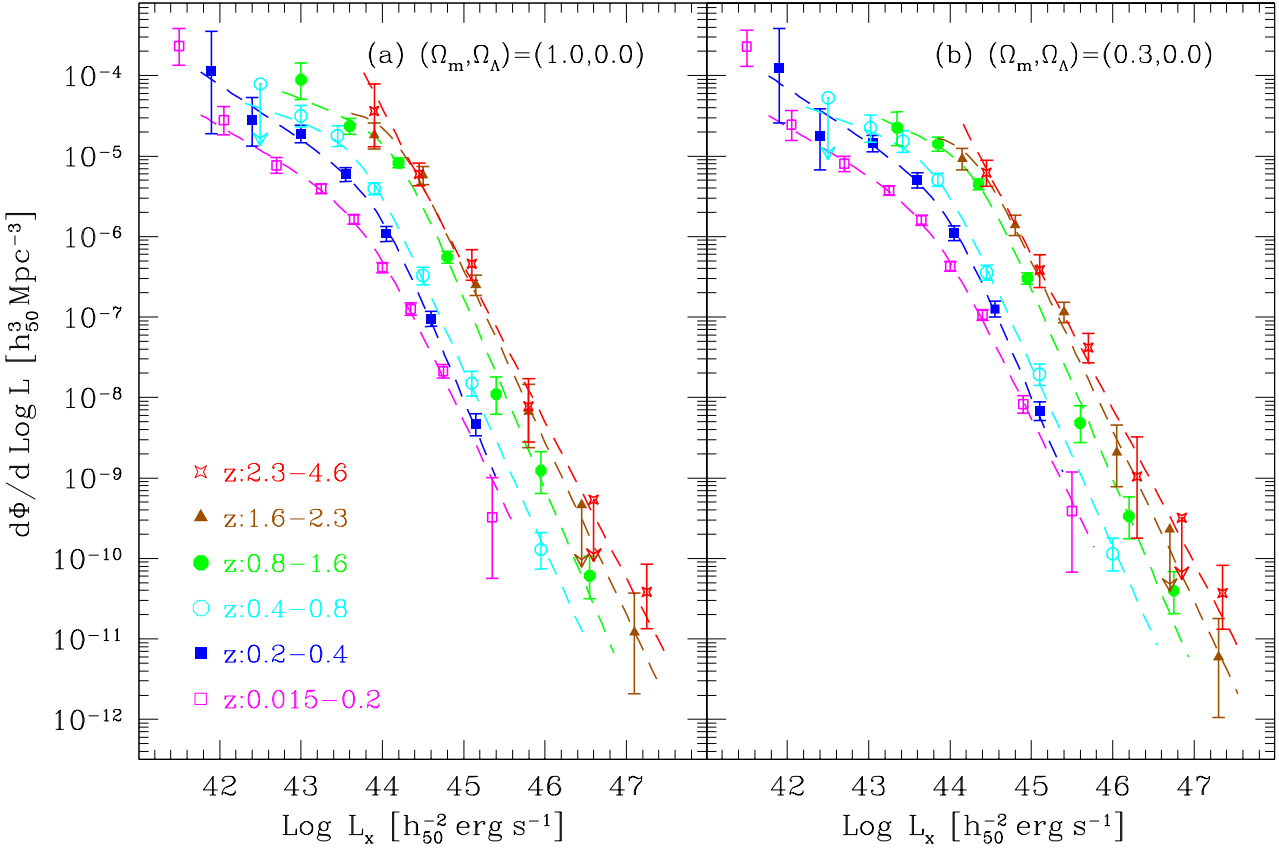}}

\caption[]{The binned SXLFs estimated by Eq.\ref{eq:est} are 
plotted with Poisson errors corresponding to the significance
range of Gaussian 1$\sigma$ from Eqs. (7) and (12) by Gehrels
(\cite{gehrels}). The data points and error estimates are more
accurate than Fig. 3 of paper I. Different symbols 
correspond to different redshift bins as indicated in the lower-left 
part of panel (a). The symbol attached to a downward arrow
indicates the 90\% upper limit (corresponding to 2.3 objects) for 
the bin with no AGN in the sample. The best-fit analytical
model for each redshift bin in the luminosity range used for the 
fit is overplotted in dashed lines.}
 
\label{fig:exlf}
\end{figure*}
    
 Using the $N^{\rm obs}/N^{\rm mdl}$ estimator,
we revised the full SXLF plot (Fig. 3 of paper I), as shown 
in Fig. \ref{fig:exlf}. Instead of connecting
the data points, we overplotted the analytical model for
each redshift bin. 

 The resulting binned SXLF are listed in Tables \ref{tab:xlf_f10},
\ref{tab:xlf_o3}, and \ref{tab:xlf_f3} for different sets of cosmological
parameters respectively. The columns of these tables are --- (1) the
redshift range of the bin; (2) the luminosity range of the bin; 
(3) the number of AGNs in the sample for the bin and the number
of non-``type 1'' AGNs as defined in Appendix A of paper I; (4) the 
number of AGNs expected from the analytical model derived in 
Sect. \ref{sec:ana}. (5) the binned SXLF estimated using Eq. \ref{eq:est} 
using the model XLF evaluated at the central point of the bin, i.e. 
$z=z_{\rm c}$ (see Table \ref{tab:fit}) and 
${\rm Log} L_{\rm x} = ({\rm Log} L_{\rm x min}+{\rm Log} L_{\rm x max})/2$,
where the subscripts min and max signify the borders of the bin
in columns (1) and (2). The upper and lower errors correspond to
Poisson errors estimated by Eqs. (7) and (12) of Gehrels (\cite{gehrels})
respectively using S=1 (corresponding to the confidence of the Gaussian
1$\sigma$. When there is no object in the bin, the Poissonian 90\% 
confidence upper-limit is given (corresponding to 2.3 objects). 
We recommend use of the values and errors under this column
when, e.g. overplotting observed SXLF values with model predictions.
\thanks{The ASCII versions of these tables with additional columns 
and separate tables with only ``type 1'' AGNs are
provided as a part of the {\em source} in the preprint archive 
of this paper (astro-ph/0101279).} 

 Figure \ref{fig:exlf} shows that the lowest-redshift, lowest-luminosity 
bin has a significant excess over the two power-law analytical expression 
(AGNs belonging to this bin have not been used for the two power-law fit), 
thus the actual underlying SXLF has a much steeper slope than that used for 
$N^{\rm mdl}$.  In order to evaluate the bias caused by this, we have made an 
$N^{\rm obs}/N^{\rm mdl}$ estimate of this particular bin using the 
{\em local} slope of $\gamma = 1.7$ instead of $\gamma \sim 0.6$ from 
the two power-law model. This gave a value about 10\% lower, thus, 
the difference is much smaller than the statistical errors for this 
bin.
   
\section{Discussion}

 We have shown the tables of observed SXLF values for a number of standard
set of cosmological parameters. These values are intended for
direct comparison with models and plotting with realistic
error bars. However, we list a number of caveats and sources of 
uncertainties, and related issues.

\begin{itemize}

\item {\em Countrate-to-flux conversion:}  For the PSPC-based observations,
 where we can limit pulse-height channels, the uncertainty in the 
 countrate-flux (in 0.5-2 [keV]) conversion is small ($\pm$ 3\% for
 photon spectral index of $\Gamma = 2.0\pm 0.7$. At the faintest
 end ($S_{\rm x14}<0.5$), where only the HRI data are available, the 
 conversion rate varies by $\pm 40\%$ for the same spectral index range.   

\item {\em Optical classification of AGNs:} Since different catalogs used
 in this analysis have different criteria for type I and type II
 AGNs, we did not show separate expressions for these two populations.
 See appendix A. of paper I for the approximate difference in behavior
 of the tentative ``type I'' sample.

\item {\em Incompleteness:} Most of the surveys used in the analysis
 are highly complete or we have selected an appropriate 
 complete subset. In case there is incompleteness, we have
 corrected for it by assuming 
 that the redshift distribution/content of the remaining sources
 are the same as the identified ones in the same flux range. 
 This assumption is not likely to be the case, considering that
 they have not been identified not because of a random cause but
 because of optical faintness and difficulty in obtaining decent
 optical spectra. The only place that this could affect significantly
 is faintest end of RDS-LH ($0.17 \le S_{\rm x14} \le 0.38$), where
 the identification completeness is $\sim 80\%$. 
 (At $S_{\rm x14}\ge 0.38$, the identification completeness
 is $\ga 95$\% in any flux range). This could affect the behavior of,
 e.g. the apparent break at the low luminosity end in 
 $1.6 \le z <2.3$.   
  
\item {\em X-ray spectra/absorption:} A serious model composer
 should be aware that the luminosity given here is for 
 $0.5-2$ [keV] in the observer frame. Thus, one should compare the
 model, with their own spectral assumptions (spectral index,
 absorption, fraction of absorbed AGNs which may depend on
 luminosity/redshift), should compute the apparent luminosity
 in the $0.5(1+z)-2(1+z)$[keV] range and compare it with the values
 listed in this paper. The latest population synthesis
 model based on absorbed and unabsorbed AGNs by Gilli et al.   
 (\cite{gilli01}) has applied this approach using
 the tables shown in this paper. 
 However, as discussed in paper I., the no K-correction case corresponds 
 to a K-correction assuming a power-law photon index of $\Gamma=2$, 
 which is the most representative spectrum for the soft X-ray sources 
 in the sample. Thus for many purposes, considering our tabulated 
 values as K-corrected SXLF would be accurate enough.  

\item {\em Large-Scale Structure:} The lowest redshift bin 
 covers 0.015 $<z<$ 0.2 and there is some concern about the 
 effect of the large-scale structure of the universe, which could be
 confused with the effect of evolution. Zucca et al. \cite{zucca97} found 
 an underdensity of galaxies in the local universe out to z$\sim$ 0.05. 
 However, it might be because of the structure within their 
 survey field of $\sim 27$ [deg$^2$] rather than that of the entire 
 space out to this redshift. The solid angle surveyed
 by RBS is $\sim$ 50\% of the sky and the fields of SA-N are scattered 
 in various directions.  In any event, our lowest redshift bin samples 
 a sufficiently large volume of space to $z= 0.2$ with a uniform redshift 
 coverage, thus it is unlikely that the calculated SXLF is
 significantly biased by the large-scale structure of the universe.

\end{itemize}

\acknowledgements
This work is based on a combination of extensive {\it ROSAT} surveys
from a number of groups. Our work is indebted to  the effort of the 
{\it ROSAT} team and the optical followup teams in producing data 
and the catalogs used in the analysis. 
In particular, we thank K. Mason, A. Schwope,
G. Zamorani, I. Appenzeller, and I. McHardy for providing us with and 
allowing us to use their data prior to 
publication of the catalogs.  TM was supported by a fellowship from the
Max-Planck-Society during his appointment at MPE. GH acknowledges 
DLR grant FKZ 50 OR 9403 5.   We thank the referee, T. Shanks, for
useful comments.

\end{document}